\documentclass[pdftex,12pt,a4paper]{article}

\usepackage[squaren]{SIunits}
\usepackage{graphicx}
\usepackage{xcolor}
\usepackage{authblk}

\begin{document}

\title{Measurement of 1.7 to 74\,MeV polarised $\gamma$ rays with the HARPO TPC}


\author[1]{Y.~Geerebaert\thanks{corresponding author: yannick.geerebaert@polytechnique.fr}}
\author[1]{Ph.~Gros\thanks{corresponding author: philippe.gros@llr.in2p3.fr}}
\author[5]{S.~Amano}
\author[2]{D.~Atti\'e}
\author[1]{D.~Bernard}
\author[1]{P.~Bruel}
\author[2]{D.~Calvet}
\author[2]{P.~Colas}
\author[6]{S.~Dat\'e}
\author[2]{A.~Delbart}
\author[1]{M.~Frotin}
\author[1]{B.~Giebels}
\author[3,4]{D.~G\"otz}
\author[5]{S.~Hashimoto}
\author[1]{D.~Horan}
\author[5]{T.~Kotaka}
\author[1]{M.~Louzir}
\author[5]{Y.~Minamiyama}
\author[5]{S.~Miyamoto}
\author[6]{H.~Ohkuma}
\author[1]{P.~Poilleux}
\author[1]{I.~Semeniouk}
\author[2]{P.~Sizun}
\author[5]{A.~Takemoto}
\author[5]{M.~Yamaguchi}
\author[1]{S.~Wang}

\affil[1]{LLR, \'Ecole Polytechnique, CNRS/IN2P3, Palaiseau, France}
\affil[2]{CEA, Irfu, CEA-Saclay, France}
\affil[3]{AIM, CEA/DSM-CNRS-Universit\'e Paris Diderot, France}
\affil[4]{IRFU/Service d'Astrophysique, CEA-Saclay, France}
\affil[5]{LASTI, University of Hy\^ogo, Japan}
\affil[6]{JASRI/SPring8, Japan}

\maketitle

\begin{abstract}
Current $\gamma$-ray telescopes based on photon conversions to
electron-positron pairs, such as Fermi, use tungsten converters. They
suffer of limited angular resolution at low energies, and their
sensitivity drops below 1\,GeV. The low multiple scattering in a
gaseous detector gives access to higher angular resolution in the
MeV-GeV range, and to the linear polarisation of the photons through
the azimuthal angle of the electron-positron pair.

HARPO is an R\&D program to characterise the operation of a TPC (Time
Projection Chamber) as a high angular-resolution and sensitivity
telescope and polarimeter for $\gamma$ rays from cosmic sources. It
represents a first step towards a future space instrument.
A 30\,cm cubic TPC demonstrator was built, and filled with 2\,bar argon-based gas. 
It was put in a polarised $\gamma$-ray beam at the NewSUBARU
accelerator in Japan in November 2014. Data were taken at different
photon energies from 1.7\,MeV to 74\,MeV, and with different
polarisation configurations. 
The electronics setup is described, with an emphasis on the trigger system.
The event reconstruction algorithm is quickly described, and preliminary 
measurements of the polarisation of 11\,\mega\electronvolt photons are shown.
\end{abstract}




\section[High angular resolution gamma-ray astronomy and polarimetry in the MeV - GeV energy range]{High angular resolution $\gamma$-ray astronomy and polarimetry in the MeV - GeV energy range}

$\gamma$-ray astronomy provides insight into understanding the
non-thermal emission of some of the most violent objects in the
Universe, such as pulsars, active galactic nuclei (AGN) and
$\gamma$-ray bursts (GRBs), and thereby understanding the
detailed nature of these objects.

Alas, between the sub-MeV and the above-GeV energy range for which
Compton telescopes ($\gamma e^- \rightarrow \gamma e^-$) and pair
telescopes ($\gamma Z \rightarrow Z e^+ e^-$) are, respectively,
highly performant, lies the MeV-GeV range over which the sensitivity
of past measurements was very limited, in particular as the
degradation of the angular resolution of pair telescopes at low energy
ruins the detection sensibility.
This hinders the observation and the understanding of GRBs, whose
spectra mostly peak in the MeV region;
it could also bias the description of the Blazar sequence.
More generally, it limits the detection of crowded regions of the
$\gamma$-ray sky such as the galactic plane to its brightest sources:
to a large extent, the MeV-GeV sensitivity gap \cite{Schoenfelder} is an angular
resolution issue.

\begin{figure} [th]
 \begin{center} 
 \includegraphics[width=0.75\linewidth]{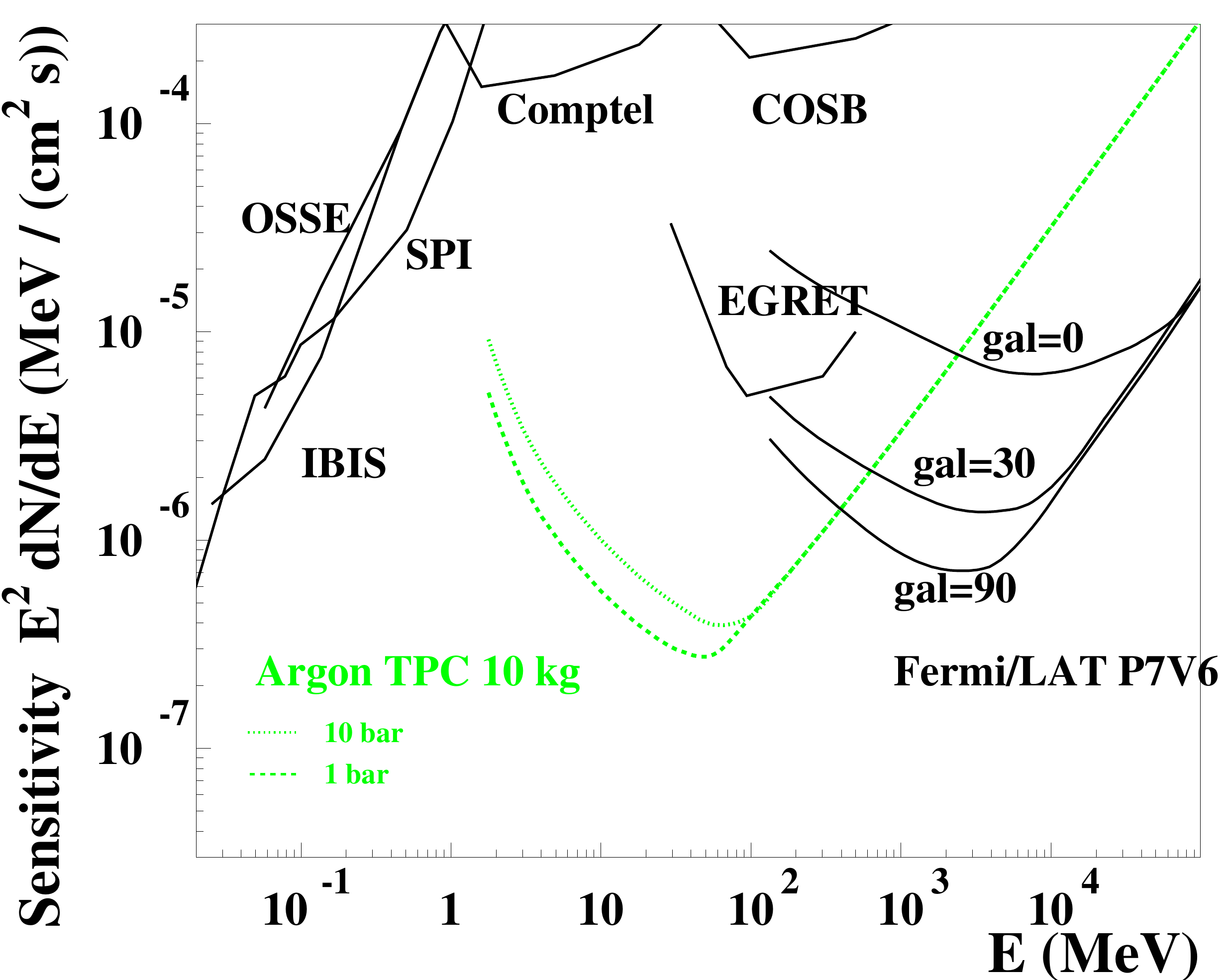}
\caption{
\label{fig:sensitivity}
 Differential sensitivity as a function of energy
 (argon-gas-based HARPO TPC, green) compared to the $90\degree$ galactic latitude
 performance of the Fermi-LAT \cite{Ackermann:2012kna} and of the
 Compton telescope COMPTEL \cite{Schoenfelder}.
 Adapted from \cite{Bernard:2012uf}.
}
\end{center}
\end{figure}

The angular resolution of pair telescopes can be improved, from the
Fermi/LAT's $\approx 5\degree$ at $100\,\mega\electronvolt$ \cite{Ackermann:2012kna} 
to $1 - 2 \degree$, by the use of pure silicon trackers, i.e. without
any tungsten converter plates \cite{ASTROGAM,PANGU,Compair}. 
An even better resolution of $\approx 0.4\degree$ can be obtained with
a gas detector such as a time-projection chamber (TPC), so that
together with the development of high-performance Compton telescope,
filling the sensitivity gap for point-like sources at a level of
$\approx 10^{-6}\,\mega\electronvolt/(\centi\meter^2 \second)$ is within
reach~(\cite{Bernard:2012uf} and Fig.~\ref{fig:sensitivity}).

Furthermore the measurement of the linear polarisation of the
emission, which is a powerful tool for understanding the characteristics of
cosmic sources at lower energies in the radio-wave to X-ray energy range,
is not available for $\gamma$-rays above $1\,\mega\electronvolt$
\cite{Mattox}.
The use of a low density converter-tracker, such as a gas detector,
enables the measurement of the polarisation fraction before 
multiple scattering ruins the azimuthal information carried by
the pair \cite{Bernard:2013jea}.

$\gamma$-ray polarimetry would provide insight into understanding the
value and turbulence of magnetic fields in the $\gamma$-ray emitting
jet structures of most $\gamma$-ray emitting sources. And, for example,
could enable us to distinguish between the leptonic and hadronic nature of the emitting
particles in blazars \cite{Zhang:2013bna}.

\section{The HARPO detector}

We have built a TPC which is using argon-based
gas mixtures in the range $1 - 5\,\bbar$ \cite{Pisa2012}. 
The drifted electrons are amplified by a hybrid amplifier
whose performance has been characterised in detail
\cite{Gros:TIPP:2014}.

We have exposed the detector to a tunable $\gamma$-ray source, using
the head-on inverse-Compton scattering of a laser beam on the 
$0.6 - 1.5\,\giga\electronvolt$ 
electron beam of the NewSUBARU storage ring (Hy\={o}go, Japan)~\cite{Horikawa2010209}.
The detector was positioned so that the photon beam would be aligned with the drift direction~$z$ of the TPC, coming from the readout side, and exiting through the cathode.

\begin{figure} [th]
 \begin{center} 
  \includegraphics[width=88mm]{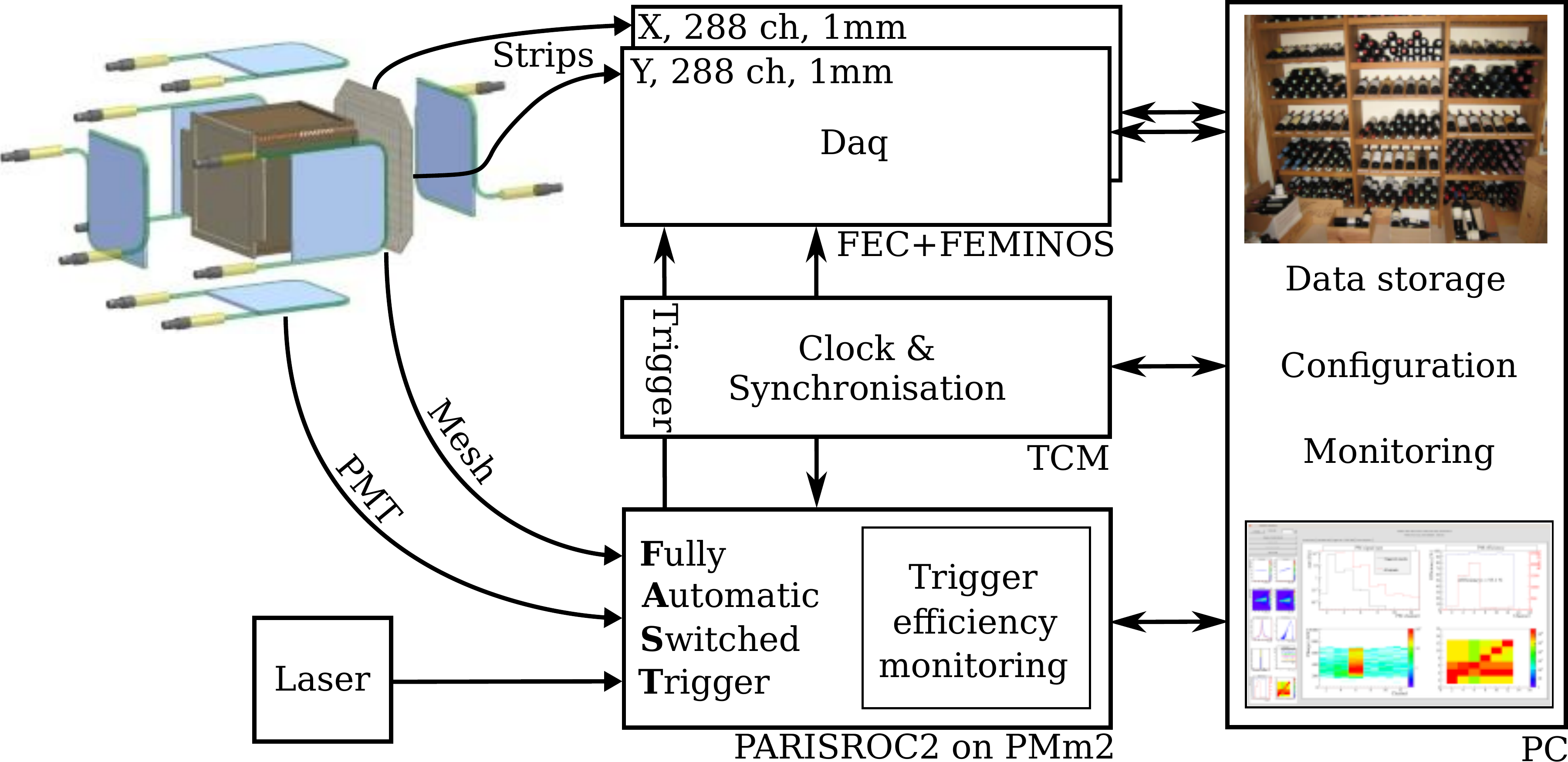}
  \caption{
   \label{fig:elec_global_view}
   Global view of HARPO electronics.
  }
 \end{center}
\end{figure}

The present detector is aimed at ground-validation tests, but
we have designed it taking into account the constraints of space
operation.

HARPO produces very fine 3D images of $\gamma$-ray conversions to
$e^+e^-$ pairs by tracking these events. This is done at a low cost
in terms of power consumption and data flow in the presence of
a large number of background noise tracks.

The TPC is a $30\centi\meter$ cubic field cage, enclosed with a copper
cathode and a readout plane anode including a hybrid multi-stage
amplification system composed of two GEM (Gas Electron Multiplier) + one MICROMEGAS 
($128\,\micro\meter$-gap bulk Micro MEsh GAseous Structure).

\begin{figure} [th]
 \begin{center} 
  \includegraphics[width=57mm]{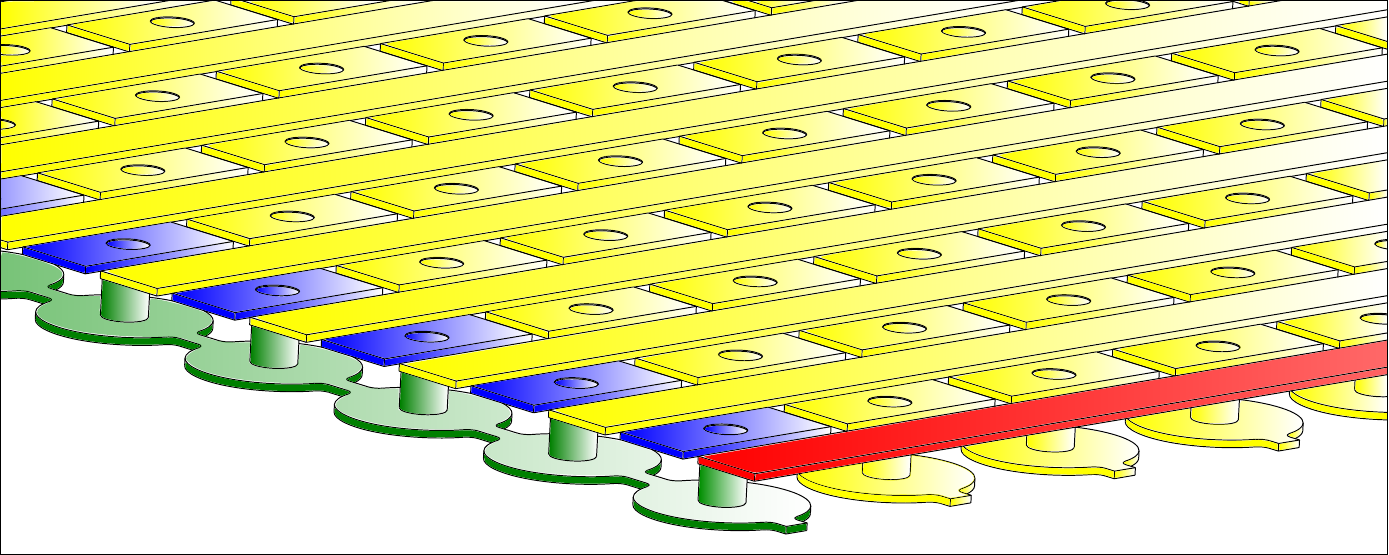}
  \hspace{2mm}
  \includegraphics[width=23mm]{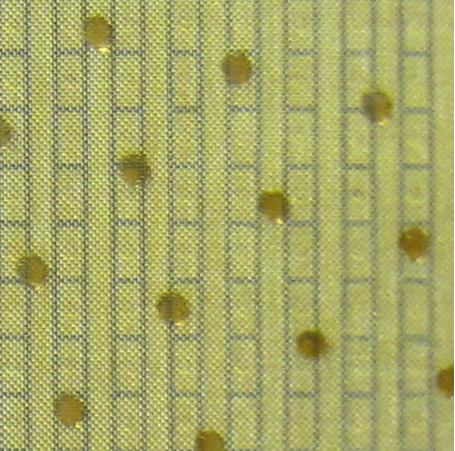}
  \caption{
   \label{fig:StripsLayout2}
   Left: Layout of readout plane of MICROMEGAS. Only the copper of the PCB-based plane is shown. One $x$ strip is coloured in red and one $y$ pseudo-strip is in blue. $y$ pseudo-strip are segmented in pads, connected together through via by an internal layer (green). Right: Top view of real PCB with pillars, keeping the mesh at a distance of $128\,\micro\meter$ of the readout plane.
  }
 \end{center}
\end{figure}

As shown in fig.~\ref{fig:StripsLayout2}, the signal
is collected by two orthogonal series of strips ($x$ \& $y$), which,
in our case, reduces the number of channels by a factor 144
compared to the equivalent pixel sensor. This reduction is only
possible if the channel occupancy is low enough to avoid
unsolvable ambiguities and comes at the cost of the need for
off-line association of each $x$ track to a track in the $y$ view. 
Then, only 576 channels ($x$ \& $y$ strips, $1\,\milli\meter$ pitch) are read out and
digitised at $33\,\mega\hertz$ (up to $100\,\mega\hertz$) by eight AFTER chips
mounted on two FEC boards. Channel data are then zero
suppressed and sent to a PC via Gigabit Ethernet by two FEMINOS
boards synchronised by one TCM board. These versatile boards
were originally developed at IRFU for the T2K and MINOS experiments~\cite{Calvet2014zva}.
To mitigate the dead time induced by readout and
digitisation ($1.6\,\milli\second$) we developed a sophisticated trigger, with
a multi-line system so as to provide real-time efficiency monitoring of each component.

\begin{figure} [th]
 \begin{center} 
  \includegraphics[width=75mm]{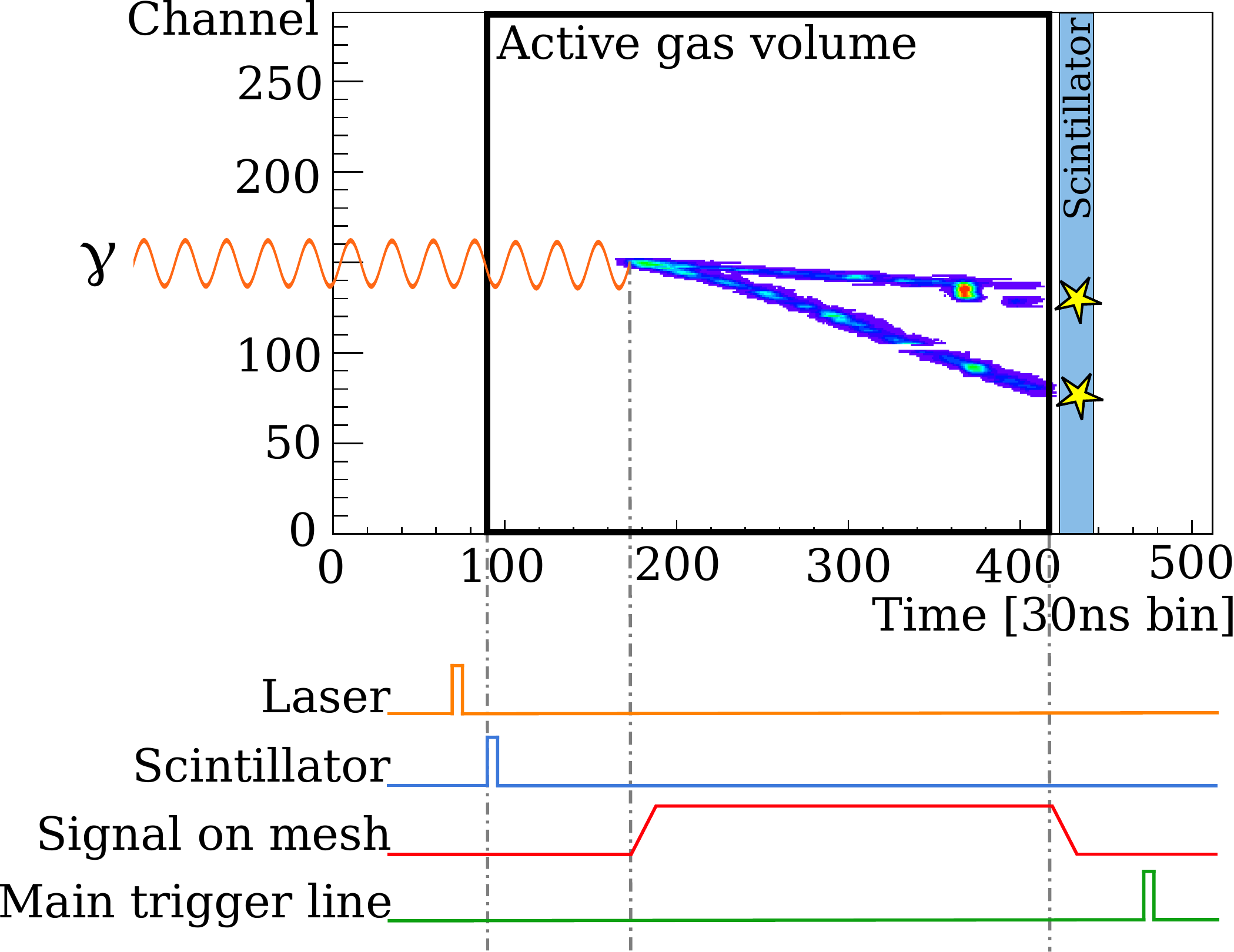}
  \caption{
   \label{fig:Timing}
   Typical timing of trigger in TPC.
  }
 \end{center}
\end{figure}

To provide presence and timing information on events, six scintillators surround the TPC.
Each scintillator is equipped with a pair of photomultiplier tubes (PMTs) which are read out by a PARISROC2 chip\cite{ConfortiDiLorenzo:2012sm} mounted on a PMM2 board\cite{Genolini:2008uc}.

The timing of the charge induced on the mesh is used for trigger. The signal, which is long with an unpredictable shape, corresponds to the time distribution of tracks in the TPC. To get a signal from the rising edge, we use a constant fraction discriminator (CFD), which shows the beginning of the signal, and therefore the position of the beginning of the track. We can measure the delay between the start of the
event and this signal to build a veto on tracks created upstream from the TPC.

The main line of the trigger selects pair-creation events which follow from the interaction of a $\gamma$ photon with the nucleus of a gas atom in the TPC. It is composed of:
\begin{itemize}
 \item a veto on upstream scintillator (which reveal an interaction before the active gas region),
 \item a signal on mesh of MICROMEGAS with a veto based on the presence of a very early signal (we reject most of the $\gamma$-rays that convert in the material of the readout plane),
 \item a signal in at least one of the five others scintillators,
 \item a laser signal, whenever available (for the pulsed laser), in coincidence with the signal in the scintillators.
\end{itemize}

This trigger suppressed the huge background rate from
the accelerator (up to $5\,\kilo\hertz$) by a rejection factor of greater than
two orders of magnitude~\cite{Wang:2015thesis} during the data-taking campaign.

\section{Analysis of data from a polarised photon beam}

We took data with the HARPO TPC in a polarised photon beam at the NewSUBARU~\cite{Wang:TPC:2015}.
The photon is aligned with the drift direction~$z$ and arrives from the readout side.
The detector was rotated around the $z$-axis to study the systematic angular effects related to the cubic geometry of the TPC.
A total of about 60~million events were taken, with 13 different photon energies from 1.7 to 74\,\mega\electronvolt, and 4 the TPC orientations.
Both polarised and unpolarised beams were used.

\begin{figure} [th]
 \begin{center} 
  \includegraphics[width=57mm]{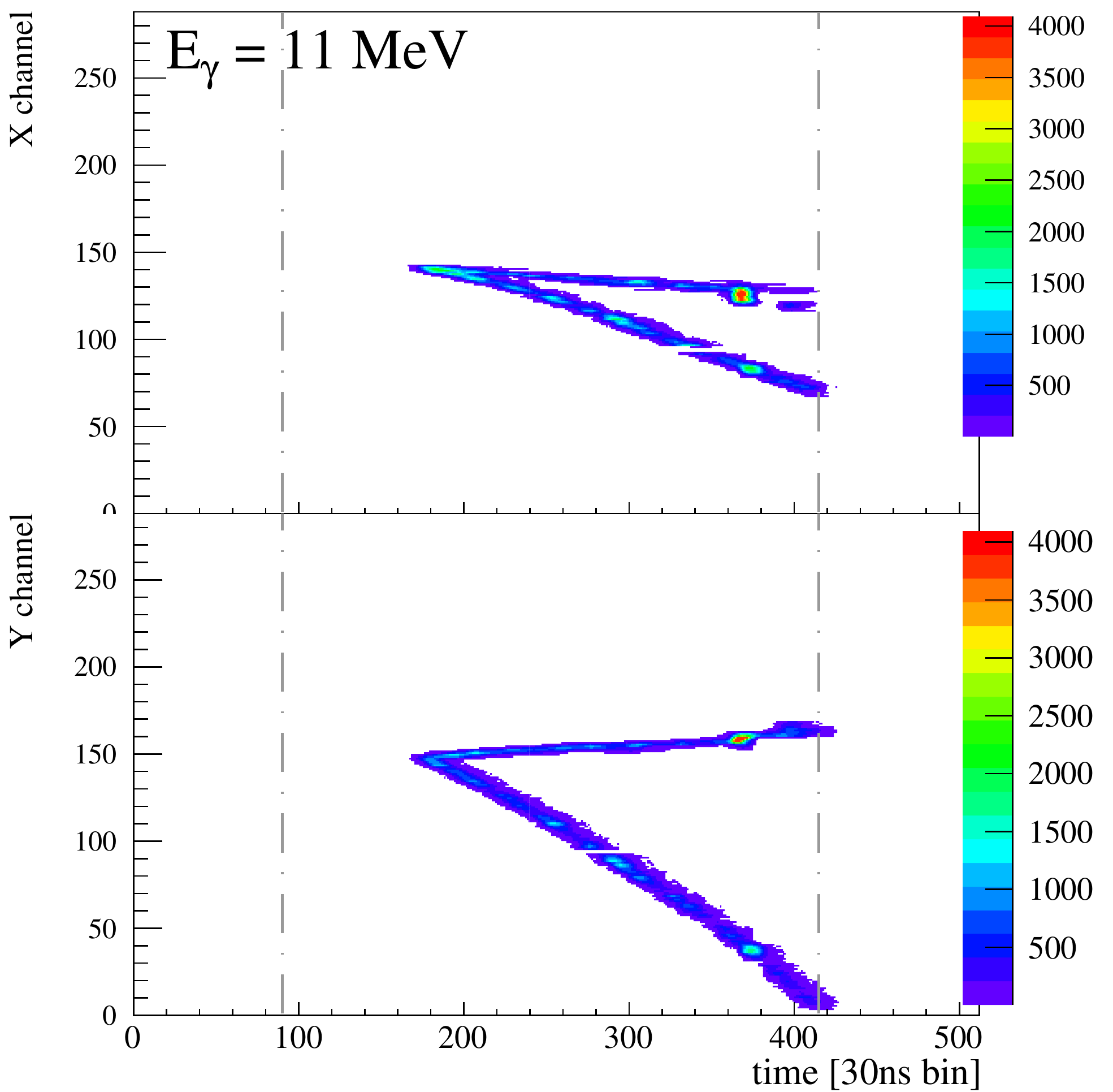}
  \caption{
   \label{fig:Event}
   Example of a raw event in the HARPO TPC.
   The two tracks from the pair conversion of a 11\,MeV photon are clearly visible.
  }
 \end{center}
\end{figure}

Figure~\ref{fig:Event} shows a pair conversion event as observed in the HARPO TPC. 
The two electron/positron tracks are visible in each of the two projections (X-Z and Y-Z).

The electron tracks were reconstructed on each projection using a closest-neighbour search based on a Kalman filter~\cite{Fruhwirth:1987fm}.
Unfortunately, since the two tracks are difficult to disentangle near the vertex, it was impossible to use the Kalman filter to estimate the track direction near the vertex.
The identified tracks were therefore fitted with a straight line.
Then, the tracks reconstructed on each projection (X-Z and Y-Z) were paired together using their charge profile as a function of the drift time (equivalent to the Z coordinate).
In this way we were able to define tracks in 3D.

For each pair of reconstructed 3D tracks, the point and distance of closest approach (POCA and DOCA) were calculated.
We selected only the track pairs where the POCA was close enough to the vertex position, as estimated from the beam geometry.
The azimuthal angle~$\omega$ for an $e^+e^-$~pair with direction $\vec{u^+}$ and $\vec{u^-}$ is defined in Fig.1 of~\cite{Wojtsekhowski2003605}.
Using the fact that our photon beam is aligned with the $z$~direction, $\omega$ is:
\begin{equation}
\omega = \arctan\left(\frac{u^-_zu^+_x-u^+_zu^-_x}{u^-_zu^+_y-u^+_zu^-_y}\right)
\end{equation}

\begin{figure} [th]
 \begin{center} 
  \includegraphics[width=57mm]{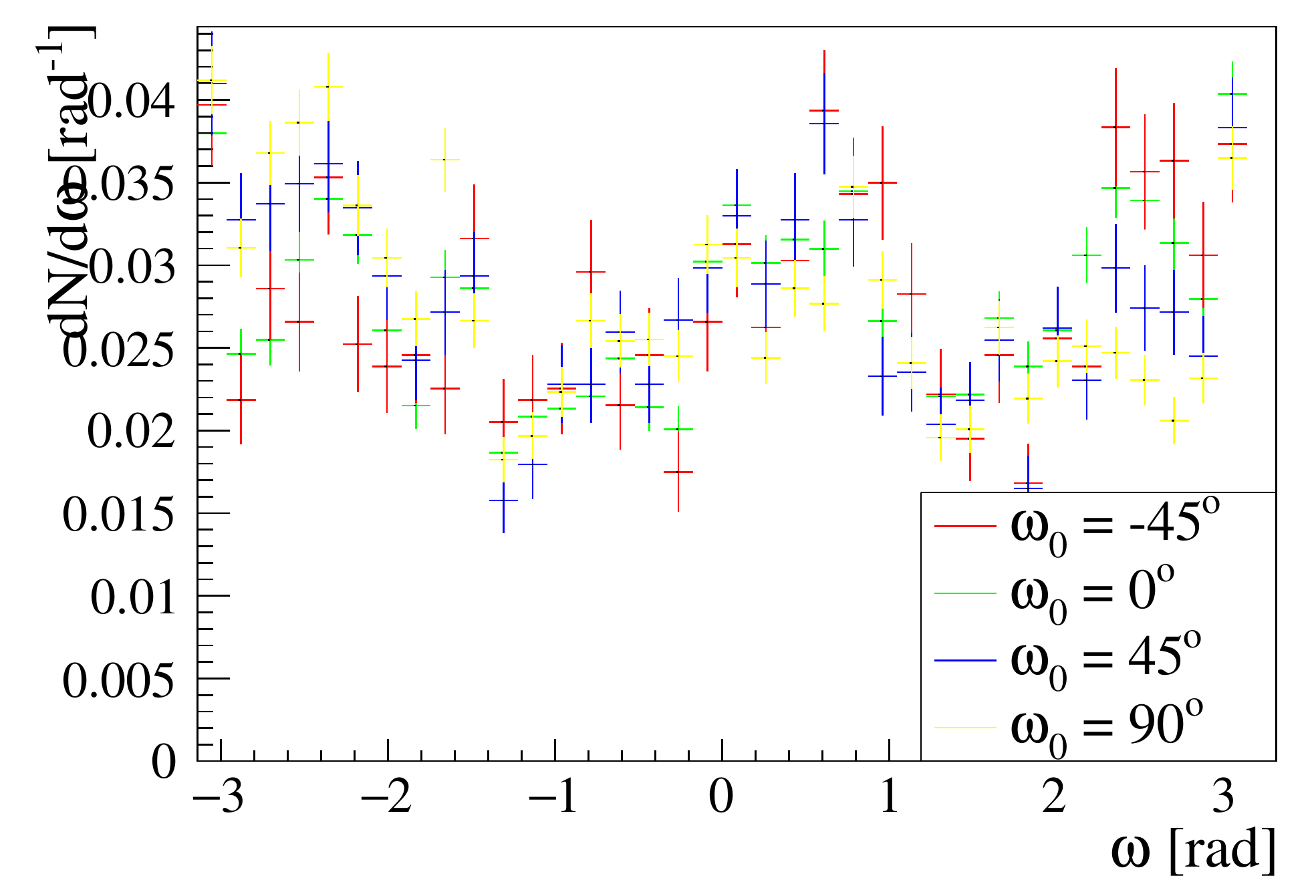}
  \caption{
   \label{fig:omega}
   Azimuthal angle $\omega$ for one configuration of the TPC with an 11\,MeV photon beam.
   Only the statistical uncertainties are shown.
   Systematic effects dominate the distribution.
  }
 \end{center}
\end{figure}

A distribution of $\omega$ is shown in Fig.~\ref{fig:omega} for several orientations of the detector with an unpolarised beam.
The distribution is dominated by systematic fluctuations due to inefficiencies of the reconstruction algorithm in some track configurations and to the cubic shape of the detector.

\begin{figure} [th]
 \begin{center} 
  \includegraphics[width=0.850\linewidth]{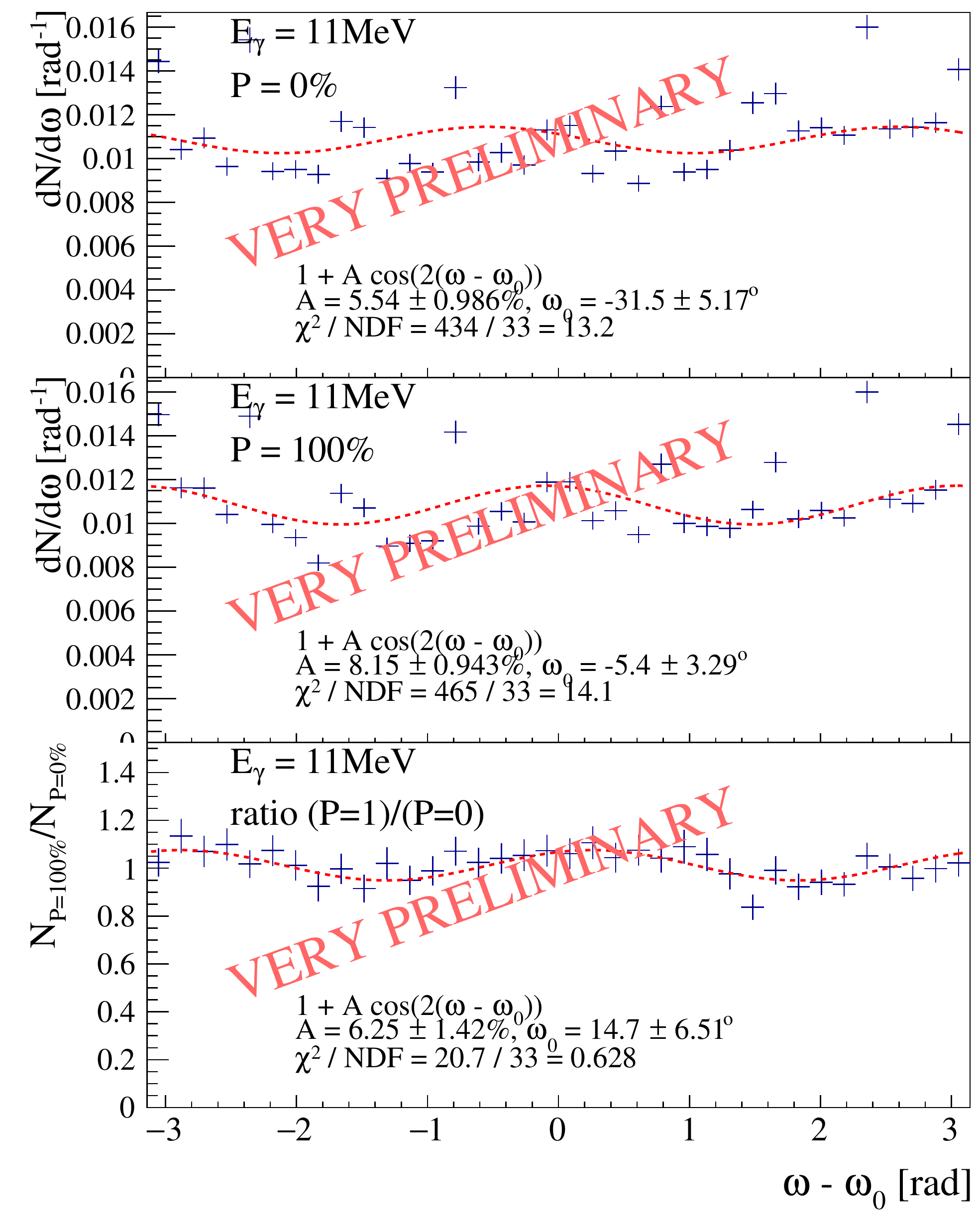}
  \caption{
   \label{fig:Pola}
   Distribution of the reconstructed azimuthal angle $\omega-\omega_{0}$ from pair conversions of 11\,MeV photons in the TPC.
   Top: unpolarised photon beam.
   Middle: fully polarised photon beam.
   Bottom: ratio of the two distributions above.
   The systematic effects are cancelled, and an effective polarisation asymmetry of 6\% is visible.
  }
 \end{center}
\end{figure}

To compensate for these systematic effects, we took data with different orientations of the TPC with regard to the photon beam.
By combining the data taken at different angles $\omega_{0}$ of the TPC we obtain the distributions of $\omega-\omega_{0}$ shown on the upper two panels of Fig.~\ref{fig:Pola}.
Most of the systematic fluctuations are averaged out, although a few remain.
From these distributions we can already see a difference between the unpolarised (top) and polarised (middle) beam data.

Finally we divide the distribution for a polarised beam, by the unpolarised one.
In this way all of the systematic effects cancel out, and only the polarisation asymmetry remains.
Figure~\ref{fig:Pola}, bottom plot, shows this asymmetry.
The measured polarisation angle $14.7\,\degree\pm6.5\,\degree$ is consistent with the expected value of zero.
The effective asymmetry, estimated at $6.2\pm1.4\%$, is significantly smaller than the theoretical value of $17\,\%$ (see Fig.\,21 in~\cite{Bernard:2013jea}).

\section{Discussion and outlook}

The HARPO TPC was successfully used to measure pair conversions of photons from a few \mega\electronvolt\ to 74\,\mega\electronvolt.
The trigger using photomultipliers and the direct signal from the micromegas mesh performed a suppression of the background by nearly two orders of magnitude, with an about $50\,\%$ selection efficiency as shown in~\cite{Wang:2015thesis}.

A first reconstruction algorithm was applied to the data to extract the azimuthal angle of the electron-positron pair.
By comparing data from the polarised and unpolarised beams, we extracted a first observation of the polarisation modulation for photons with 11\,\mega\electronvolt\ energy and higher.
The reconstruction efficiency is still low, and many systematic effects remain to be understood.
These results show that photon polarimetry in the pair-production regime can be performed with a TPC.
A more appropriate reconstruction is being developed to increase the efficiency and the resolution.

These preliminary results demonstrate that the design of a space TPC is viable.
Sophisticated hardware and software allows a reduction of the number of channels by several orders of magnitude.
The next step will be the design of a balloon-borne TPC.
It will be used to validate the trigger which is a key component for a successful space telescope.
The trigger will have to extract $\sim10\,\hertz$ photon conversion signal from about $\sim5000\,\hertz$ single-track background.

\section{Acknowledgements}

This work is funded by the French National Research Agency
(ANR-13-BS05-0002).

\bibliographystyle{elsarticle-num}
\bibliography{mybibfile}

\begin{thebibliography}{10}
\expandafter\ifx\csname url\endcsname\relax
  \def\url#1{\texttt{#1}}\fi
\expandafter\ifx\csname urlprefix\endcsname\relax\def\urlprefix{URL }\fi
\expandafter\ifx\csname href\endcsname\relax
  \def\href#1#2{#2} \def\path#1{#1}\fi

\bibitem{Schoenfelder}
{V. Schoenfelder}, Lessons learnt from {COMPTEL} for future telescopes, New
  Astronomy Reviews 48 (2004) 193--198.
\newblock \href {http://dx.doi.org/doi:10.1016/j.newar.2003.11.027}
  {\path{doi:doi:10.1016/j.newar.2003.11.027}}.

\bibitem{Ackermann:2012kna}
M.~Ackermann, et~al., {The Fermi Large Area Telescope On Orbit: Event
  Classification, Instrument Response Functions, and Calibration}, Astrophys.\
  J.\ Suppl.\ 203 (2012) 4.
\newblock \href {http://dx.doi.org/dx.doi.org/10.1088/0067-0049/203/1/4}
  {\path{doi:dx.doi.org/10.1088/0067-0049/203/1/4}}.

\bibitem{ASTROGAM}
{M. Tavani, V. Tatischeff}, et~al., {The ASTROGAM gamma-ray space mission; A
  sensitive observatory for the MeV - GeV domain}In preparation.

\bibitem{PANGU}
{Xin Wu, Meng Su}, et~al., {PANGU: A High Resolution Gamma-ray Space
  Telescope}, Proc.SPIE Int.Soc.Opt.Eng. 9144 (2014) 91440F.
\newblock \href {http://dx.doi.org/dx.doi.org/10.1117/12.2057251}
  {\path{doi:dx.doi.org/10.1117/12.2057251}}.

\bibitem{Compair}
A.~Moiseev, et~al., {Compton-Pair Production Space Telescope (ComPair) for MeV
  Gamma-ray Astronomy}, ArXiv e-prints\href {http://arxiv.org/abs/1508.07349}
  {\path{arXiv:1508.07349}}.

\bibitem{Bernard:2012uf}
D.~Bernard, {TPC in gamma-ray astronomy above pair-creation threshold}, Nucl.\
  Instrum.\ Meth.\ A 701 (2013) 225.

\bibitem{Mattox}
J.~R. Mattox, Analysis of the {COS B} data for evidence of linear polarization
  of {VELA} pulsar gamma rays, Astrophysical Journal 363 (1990) 270--273.
\newblock \href {http://dx.doi.org/dx.doi.org/10.1086/169338}
  {\path{doi:dx.doi.org/10.1086/169338}}.

\bibitem{Bernard:2013jea}
D.~Bernard, Polarimetry of cosmic gamma-ray sources above $e^+ e^-$ pair
  creation threshold, Nucl.\ Instrum.\ Meth.\ A 729 (2013) 765.
\newblock \href {http://dx.doi.org/dx.doi.org/10.1016/j.nima.2012.11.023}
  {\path{doi:dx.doi.org/10.1016/j.nima.2012.11.023}}.

\bibitem{Zhang:2013bna}
H.~Zhang, M.~B{\"o}ttcher, {X-Ray and Gamma-Ray Polarization in Leptonic and
  Hadronic Jet Models of Blazars}, Astrophys.\ J.\ 774 (2013) 18.

\bibitem{Pisa2012}
D.~Bernard, {HARPO - A gaseous TPC for high angular resolution $\gamma$-ray
  astronomy and polarimetry from the MeV to the TeV}, Nucl.\ Instrum.\ Meth.\ A
  718 (2013) 395--399.
\newblock \href {http://dx.doi.org/dx.doi.org/10.1016/j.nima.2012.10.054}
  {\path{doi:dx.doi.org/10.1016/j.nima.2012.10.054}}.

\bibitem{Gros:TIPP:2014}
P.~{Gros}, et~al., {HARPO - TPC for High Energy Astrophysics and Polarimetry
  from the MeV to the GeV}, Proceedings of Science TIPP2014 (2014) 133.

\bibitem{Horikawa2010209}
K.~Horikawa, S.~Miyamoto, S.~Amano, T.~Mochizuki,
  \href{http://www.sciencedirect.com/science/article/pii/S0168900210005449}{Measurements
  for the energy and flux of laser compton scattering γ-ray photons generated
  in an electron storage ring: Newsubaru}, Nuclear Instruments and Methods in
  Physics Research Section A: Accelerators, Spectrometers, Detectors and
  Associated Equipment 618~(1 - 3) (2010) 209 -- 215.
\newblock \href
  {http://dx.doi.org/http://dx.doi.org/10.1016/j.nima.2010.02.259}
  {\path{doi:http://dx.doi.org/10.1016/j.nima.2010.02.259}}.
\newline\urlprefix\url{http://www.sciencedirect.com/science/article/pii/S0168900210005449}

\bibitem{Calvet2014zva}
D.~Calvet, {A Versatile Readout System for Small to Medium Scale Gaseous and
  Silicon Detectors}, IEEE Trans. Nucl. Sci. 61~(1) (2014) 675--682.
\newblock \href {http://dx.doi.org/10.1109/TNS.2014.2299312}
  {\path{doi:10.1109/TNS.2014.2299312}}.

\bibitem{ConfortiDiLorenzo:2012sm}
S.~Conforti Di~Lorenzo, et~al., {PARISROC, an autonomous front-end ASIC for
  triggerless acquisition in next generation neutrino experiments}, Nucl.
  Instrum. Meth. A695 (2012) 373--378.
\newblock \href {http://dx.doi.org/10.1016/j.nima.2011.11.028}
  {\path{doi:10.1016/j.nima.2011.11.028}}.

\bibitem{Genolini:2008uc}
B.~Genolini, et~al., {PMm2: Large photomultipliers and innovative electronics
  for the next-generation neutrino experiments}, Nucl. Instrum. Meth. A610
  (2009) 249--252.
\newblock \href {http://arxiv.org/abs/0811.2681} {\path{arXiv:0811.2681}},
  \href {http://dx.doi.org/10.1016/j.nima.2009.05.135}
  {\path{doi:10.1016/j.nima.2009.05.135}}.

\bibitem{Wang:2015thesis}
S.~Wang, {\'Etude d'une TPC, cible active pour la polarim\'etrie et
  l'astronomie gamma par cr\'eation de paire dans HARPO}, Ph.D. thesis, \'Ecole
  Polytechnique, in French (9 2015).

\bibitem{Wang:TPC:2015}
S.~{Wang}, et~al., {HARPO: a TPC concept for $\gamma$-ray polarimetry with high
  angular resolution in the MeV-GeV range}, J. Phys. Conf. Ser. 650 (2015)
  012016.

\bibitem{Fruhwirth:1987fm}
R.~Fruhwirth, {Application of Kalman filtering to track and vertex fitting},
  Nucl. Instrum. Meth. A262 (1987) 444--450.
\newblock \href {http://dx.doi.org/10.1016/0168-9002(87)90887-4}
  {\path{doi:10.1016/0168-9002(87)90887-4}}.

\bibitem{Wojtsekhowski2003605}
B.~Wojtsekhowski, D.~Tedeschi, B.~Vlahovic,
  \href{http://www.sciencedirect.com/science/article/pii/S0168900203023143}{A
  pair polarimeter for linearly polarized high-energy photons}, Nucl.\
  Instrum.\ Meth.\ A 515~(3) (2003) 605 -- 613.
\newblock \href
  {http://dx.doi.org/http://dx.doi.org/10.1016/j.nima.2003.07.009}
  {\path{doi:http://dx.doi.org/10.1016/j.nima.2003.07.009}}.
\newline\urlprefix\url{http://www.sciencedirect.com/science/article/pii/S0168900203023143}

\end{thebibliography}

\end{document}